\documentclass[conference]{IEEEtran}
\usepackage{amssymb}
\usepackage{enumerate}
\usepackage{pifont}

\usepackage{gensymb}
\usepackage{amsmath}
\usepackage{array}
\usepackage{booktabs}
\usepackage{multirow}

\usepackage{rotating}
\usepackage{epstopdf}
\usepackage{threeparttable} 

\usepackage{color}

\usepackage{algorithmicx}
\usepackage[ruled]{algorithm}
\usepackage[noend]{algpseudocode}
\vspace{-0.2cm}%
\alglanguage{pseudocode}
\usepackage{amsthm}

\ifCLASSINFOpdf
\else
\fi

\begin{document}

\title{Exploiting PUF Models for Error Free Response Generation}

	\author{
		\IEEEauthorblockN{Yansong Gao\IEEEauthorrefmark{2}\IEEEauthorrefmark{3}, Hua Ma\IEEEauthorrefmark{2}, Gefei Li\IEEEauthorrefmark{2}, Shaza Zeitouni\IEEEauthorrefmark{4}, Said F.~Al-Sarawi\IEEEauthorrefmark{3},~Derek~Abbott\IEEEauthorrefmark{3},\\ Ahmad-Reza Sadeghi\IEEEauthorrefmark{4} and Damith C.~Ranasinghe\IEEEauthorrefmark{2}}
		\IEEEauthorblockA{\IEEEauthorrefmark{3}School of Electrical and Electronic Engineering, The University of Adelaide, SA 5005, Australia
			\\\{yansong.gao, said.alsarawi, derek.abbott\}@adelaide.edu.au}
		\IEEEauthorblockA{\IEEEauthorrefmark{2}Auto-ID Labs, School of Computer Science, The University of Adelaide, SA 5005, Australia
			\\ \{mary.ma, gefei.li, damith.ranasinghe\}@adelaide.edu.au}
		    \IEEEauthorblockA{\IEEEauthorrefmark{4}System Security Lab, Technische Universitat Darmstadt, Darmstadt 64289, Germany
		    	\\shaza.zeitouni@trust.cased.de; ahmad.sadeghi@trust.tu-darmstadt.de}
	}
	



\maketitle

\begin{abstract}
Physical unclonable functions (PUF) extract secrets from randomness inherent in manufacturing processes. PUFs are utilized for basic cryptographic tasks such as authentication and key generation, and more recently, to realize key exchange and bit commitment requiring a large number of error free responses from a strong PUF. We propose an approach to eliminate the need to implement expensive on-chip error correction logic implementation and the associated helper data storage to reconcile naturally noisy PUF responses. In particular, we exploit a statistical model of an Arbiter PUF (APUF) constructed under the nominal operating condition during the challenge response enrollment phase by a trusted party to judiciously select challenges that yield error-free responses even across a wide operating conditions, specifically, a $ \pm 20\% $ supply voltage variation and a $ 40\celsius $ temperature variation. We validate our approach using measurements from two APUF datasets. Experimental results indicate that large number of error-free responses can be generated on demand under worst-case when PUF response error rate is up to 16.68\%.
\end{abstract}


\IEEEpeerreviewmaketitle

\section{Introduction}
Physical unclonable functions (PUFs) are low-cost hardware security primitives and can be seamlessly integrated with devices during the design and fabrication processes~\cite{suh2007physical,devadas2008design,hussain2016built}, to act as trust anchors. The PUF, in essence, extracts secrets from the inevitable process variations. Hence, in reality, identical PUF instances cannot be forged, not even by the same manufacturer. 
There exists a number of PUF structures~\cite{gassend2002controlled,lim2005extracting,maiti2012robust,holcomb2009power}. Among them, time-delay based PUFs~\cite{lim2005extracting,suh2007physical}, in particular, the Arbiter PUF (APUF) and its variants such as XOR-APUFs, is one of popular silicon PUF construction considering its compact structure and the large challenge-response pair (CRP) space characterizing of a strong PUF~\cite{ruhrmair2013puf}. 

Prominent PUF applications include authentication and cryptographic key generation~\cite{suh2007physical}. In recent years, use of PUFs in more advanced cryptographic protocols such as key exchange, oblivious transfer, bit commitment and muti-party computation~\cite{ruhrmair2013pufs,van2014protocol}, where a strong PUF is always required, has been investigated. Although authentication is able to tolerate the noisy PUF responses, key generation applications and key exchange, especially recent advanced cryptographic protocols~\cite{ruhrmair2013pufs}, require large number of error free responses.

Stabilizing PUF response is usually left to the on-chip error correcting logic assisted by the associated priori computed helper data as illustrated in Fig.~\ref{fig:keygeneration}. Such a scheme maybe adequate for a single or a small number of key extractions such as from SRAM PUFs~\cite{holcomb2009power}, but becomes a limitation when a large number of keys are necessary. We propose exploiting PUF models, originally used attack PUFs by creating PUF functional copies~\cite{ruhrmair2013puf}, to enable the determination of a randomly selected challenge to generate error-free response under all operating conditions in the absence of the APUF. The developed approach is able to: i) eschew expensive error correction logic on PUF embedded security modules; ii) overcome the burden of exhaustive characterization of huge number of CRPs, especially when large number of keys extracted from strong PUFs are necessary; iii) eliminate computation of helper data and subsequent on-chip storage or off-chip storage and transfer. We summarize our contributions below:
\begin{enumerate}
\item We exploit statistical delay models of APUFs to evaluate the response reliability of a random challenge to select challenges that generate error-free responses under all operating conditions. 
\item We have validated our approach based on two APUF datasets. First, we build synthetic APUFs using real-world RO-PUF~\cite{maiti2012robust} frequency measurements to acquire arbitrary number of CRPs for performing extensive evaluations. Second, we use 64,000 CRPs collected across eight APUFs implemented on eight FPGAs. Our results show that no erroneous bits occur when selected responses are re-evaluated across a wide range of operating conditions. In addition, the selected challenges still exhibit the randomness characteristic expected from the corresponding responses.
\item We show that our error-free response generation method is 'cost-free' to the PUF integrated device without any area or power overhead. Building a model takes less than fifteen seconds including collecting a small number, 10,000, of CRPs and the subsequent delay-time statistical characterization of each delay segment at each stage of an APUF, while the reliable challenge selection using the model can be performed on demand.
\end{enumerate}
Next section introduces related work. Section~\ref{Sec:APUFModel} describes the approach to build an accurate statistical delay model of an APUF, and subsequently describes an algorithm to filter challenges that produce highly reliable response bits. Section~\ref{Sec:ExperimentalSysthesis} validates the proposed approach based on two APUF datasets. Section~\ref{Sec:Conclusion} concludes this paper.
\section{Related Work}\label{Sec:relatedwork}
\begin{figure}
	\centering
	\includegraphics[trim=0 0 0 0,clip,width=0.25\textwidth]{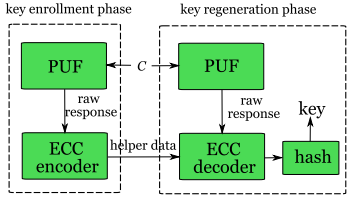}
	\caption{Generalized ECC based PUF key generation scheme.}
	\label{fig:keygeneration}
\end{figure}
Reconciling PUF response errors is usually carried out by employing error correction codes (ECC). Thus the ECC based generalized PUF key generations have two phases as illustrated in Fig.~\ref{fig:keygeneration}: (1) The key enrollment phase computes helper data that will be used in (2) the key regeneration phase to reconcile errors of the reproduced response and retrieve the enrolled key. 

This scheme is efficient to guarantee a small number of error-free responses to derive a single cryptographic key such as from a SRAM PUF \cite{holcomb2009power}, but it becomes expensive when multiple number of keys are necessary to realize: i) revocation of keys; ii) session key exchange; iii) controlled PUF constructions~\cite{gassend2002controlled,gassend2008controlled}. These applications all prefer employing a strong PUF that is able to provide a large number of error-free responses. Notably, the associated helper data leaks information that may be exploited to attack PUFs~\cite{becker2015pitfalls}.


There is a concurrent but independent investigation of selecting reliable responses of delay-based PUFs~\cite{xuusing} using simulated data, where the effect of explicit PUF operational conditions, such as operating voltage and temperature, on the selection strategy remains to be investigated. Based on measured data, we extensively evaluate our error-free response generation approach across a wide range of operating conditions. 
\section{Error Free Response Selection}\label{Sec:APUFModel}
We assume that a trusted party---the server---has one-time access to the underlying APUFs acting as a PUF building block within the controlled PUF or PUF embedded cryptographic key generators. The PUF building block can also be XOR-APUFs rather than basic APUFs. The server securely stores the derived statistical model of each APUF and destroys the one-time access, eg., fusing one-time programmable wires. The challenge response interface can be protected. For example, in a controlled PUF construction, it is achieved in a way of $ g(x)={h_2}(h_1(x),f(h_1(x))) $, where $ h $ is a hash function, $ x $ is the input, and $ f(\cdot) $ is the PUF. 

We employ a delay time adaptive characterization technique following \cite{xu2015adaptive} to build up a statistical model of an APUF that is able to accurately estimate the internal delay times of the APUF to eschew physically characterizations. In~\cite{xu2015adaptive}, the authors employ this adaptive characterization to realize an impersonation of an APUF from a physical means rather than a software means. During the key extraction phase, filtered reliable challenges are issued by the server and applied to the PUF to obtain reliable response bits that can be directly utilized to derive keys without error correction. 
Next, we detail the principles of generating error free responses when wire delay times for all stages of the APUF are estimated.
\begin{figure}
	\centering
	\includegraphics[trim=0 0 0 0,clip,width=0.40\textwidth]{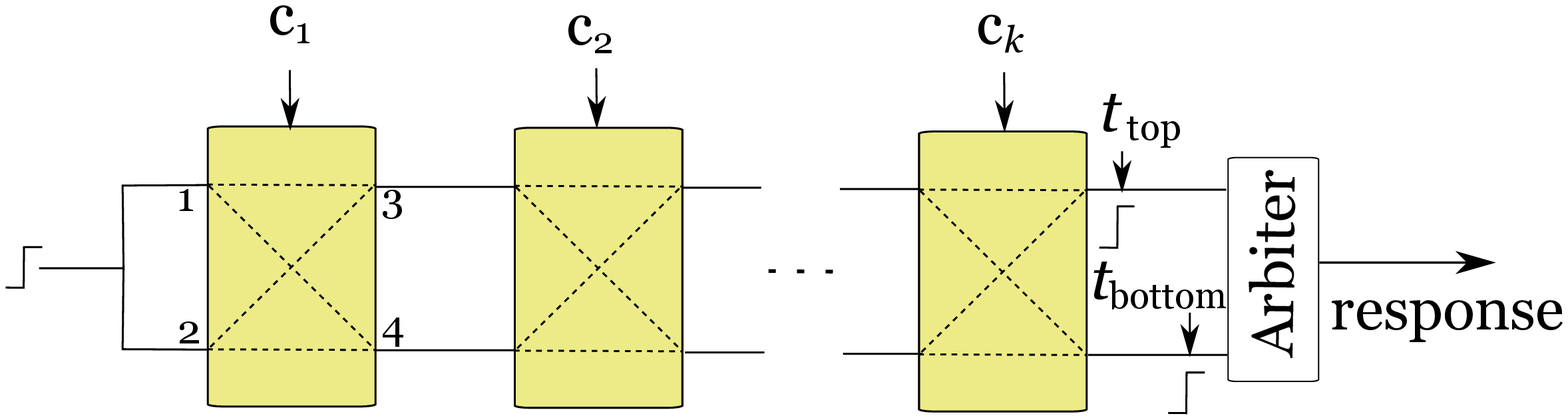}
	\caption{Structure of an arbiter PUF (APUF). Signal propagation delays of the top path $ t_{\rm top} $ and the bottom path $ t_{\rm bottom} $ in the arbiter input determines the response bits.}
	\label{fig:APUF}
\end{figure}
\subsection{Arbiter PUF}
As depicted in Fig.~\ref{fig:APUF}, an APUF consists of $k$ stages of two 2-input multiplexers, or any other unit forming two theoretically symmetrical, but practically asymmetrical, signal paths due to inherent randomness in the fabrication processes. To generate a response bit, an active pulse is fed as an input to the first stage, while the selection bit of $ c_i $, $ i\in\{1,...,k\} $ determines the signal path to the next stage. For example, if $ c_i=1 $, two signals propagate from 1 to 3 and from 2 to 4, respectively, in other words, these two signals pass through the $ i$-th stage of the APUF without crossing. Conversely, if $ c_i=0 $, they propagate from 1 to 4 and from 2 to 3 concurrently, these two paths are referred to as cross paths. We refer to the delay time from 1 to 3 of $ i_{\rm th} $ stage as $ t_{\rm 13}^{i} $, and the corresponding delays through the others as $ t_{14}^i, t_{23}^i$, and $t_{24}^i $. At the circuit level, each stage of the APUF can be implemented by a multiplexer. At the end of the cascaded multiplexers, an arbiter, which can be implemented by a latch, determines whether the top or bottom signal arrives first and hence results in a logic `0' or `1', accordingly, to yield a 1-bit response.
\begin{figure}[h]
	\centering
	\includegraphics[trim=0 0 0 0,clip,width=0.30\textwidth]{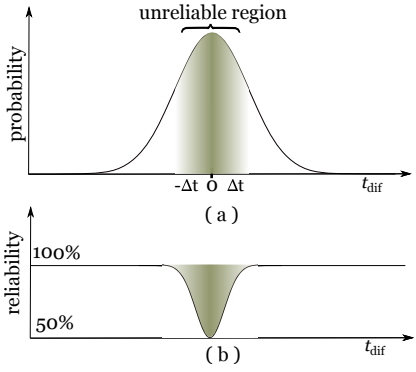}
	\caption{(a) Delay time difference $ t_{\rm dif} $ distribution, where $ t_{\rm dif}=t_{\rm top}-t_{\rm bottom} $, $ t_{\rm top} $ and $ t_{\rm bottom} $ are the signal propagation delay times of the input signal at the top path and the bottom path respectively. (b) Response bit reliability determined by $ t_{\rm dif} $ to a given challenge. 
	}
	\label{fig:TimeDelayDifferenceDist}
\end{figure}
\subsection{Response Reliability}
\begin{figure}[h]
	\centering
	\includegraphics[trim=0 0 0 0,clip,width=0.40\textwidth]{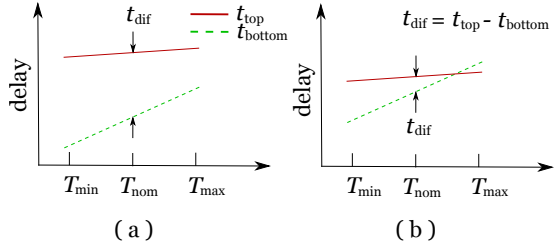}
	\caption{Delay time as a function of the temperature. (a) Challenge always reproduces the response with 100\% reliability within the temperature range from $ T_{\rm min}$ to $T _{\rm max}$. (b) Challenge results in an unreliable response bit.}
	\label{fig:delay}
\end{figure}

More specifically, the delay time difference $ t_{\rm dif}$ between the delay time of the top path $ t_{\rm top} $ and the delay time of the bottom path $ t_{\rm bottom} $ determines the response bit (output) at the arbiter and also its reliability, as depicted in Fig.~\ref{fig:TimeDelayDifferenceDist}. For a randomly chosen challenge, if the corresponding delay time difference $ t_{\rm dif}>0$, then a response of `0' is produced, and vice versa. Further, if $ |t_{\rm dif}|\le |\Delta t|$, $ \Delta t $ being a delay discrimination threshold, the delay time difference $ t_{\rm dif}$ falls into the unreliable region as illustrated in Fig.~\ref{fig:TimeDelayDifferenceDist}; then the corresponding response bit has a high probability of being nondeterministic during repeated regeneration attempts, especially under varying environmental conditions. Conversely, if $ |t_{\rm dif}|> |\Delta t| $ , then a highly reliable response bit can be expected. Further, if $ |\Delta t| $ is far away from zero, we can expect that a response bit without error will be generated. This relies on the fact that responses to those challenges resulting in a large $ |t_{\rm dif}| $ are able to tolerate a greater degree of environmental variations, eg., fluctuation of supply voltage, and temperature, as further depicted in Fig.~\ref{fig:delay}. For example, considering that the $ t_{\rm top} $ and $ t_{\rm bottom} $ have different simplified linear temperature co-efficients. A large $ t_{\rm dif} $ guarantees no intersection between  $ t_{\rm top} $ and $ t_{\rm bottom} $, while an intersection eventually results in a flipped response---erroneous response when the temperature deviates. Similar observation has shown in~\cite{suh2007physical}, where a pair of ring oscillators (ROs) exhibiting a large frequency difference ensures a reliable response bits regenerated from a ROPUF. 

If $ t_{13}^i, t_{14}^i, t_{23}^i$, and $t_{24}^i $ can be accurately measured, then it is easy to determine challenges capable of generating highly reliable response bits by evaluating $ |t_{\rm dif}| $. Concurrently, the response bit value, being `0' or `1', can also be judiciously determined. Physical measurement of each delay segment is hard, if not impossible, in practice. It may require probing the die which may damage the circuit or alter the delay, or other on-chip complicated peripheral circuits that are always unavailable for resource-constraint devices~\cite{majzoobi2010rapid}. Next, we introduce a method following \cite{xu2015adaptive} for obtaining a statistical model to capture delay times of internal wires for all $ k $ stages of an APUF.
\subsection{Modeling an APUF}
The goal of a statistical model is to accurately characterize individual delay times of $ t_{13}^i, t_{14}^i, t_{23}^i$, and $t_{24}^i $. More specifically, if the probability of $ t_{13}^i>t_{24}^i $, $  P(t_{13}^i>t_{24}^i) $, and probability of $ t_{14}^i>t_{23}^i $, $ P(t_{14}^i>t_{23}^i) $, can be estimated precisely, then $  P(t_{13}^i>t_{24}^i) $  and $ P(t_{14}^i>t_{23}^i) $ can be used to replace the  delay time difference of $ t_{13}^i-t_{24}^i $ and $ t_{14}^i-t_{23}^i $ at the $ i $-th stage of an APUF. In other words, we need to statistically estimate the following four probabilities in (\ref{Eq:3}) after collecting a number of CRPs.
\begin{align}\label{Eq:3}
\begin{split}
P_{13}^i=P(t_{13}^i \text{ is longer than } t_{24}^i \text{ at {\it i}-th stage})\\
P_{24}^i=P(t_{24}^i \text{ is longer than } t_{13}^i \text{ at {\it i}-th stage})\\
P_{14}^i=P(t_{14}^i \text{ is longer than } t_{23}^i \text{ at {\it i}-th stage})\\
P_{23}^i=P(t_{23}^i \text{ is longer than } t_{14}^i \text{ at {\it i}-th stage})
\end{split}
\end{align}
Given the probabilistic formula of delays, there exists two relations $P_{13}^i +P_{24}^i=1$ and $P_{14}^i+P_{23}^i=1$ for $i \in \{1,2,...,k\}$.  

The $ P_{13}^i$, $ P_{24}^i$, $ P_{14}^i$, $ P_{23}^i$ for all stages are estimated based on an adaptive characterization technique. We refer readers to Xu {\it et .al}'s work for detailed implementations~\cite{xu2015adaptive}. Consequently, a statistical delay model of an APUF is obtained when $ P_{13}^i$, $ P_{24}^i$, $ P_{14}^i$, $ P_{23}^i$ for all stages are estimated. As there is a linear relation between the delay time, eg., $ t_{12} $, and its corresponding probability, eg., $ P_{12} $~\cite{xu2015adaptive}, we can use the corresponding probability to replace the delay time and consequently to predict responses and concurrently their reliability for unseen challenges according to $ t_{\rm dif} $ that is the summation of delay difference for all stages.
\subsection{Determining Reliable Challenges}
Filtering reliable response bits is straightforward and simple, once the statistical model is obtained, and it is a one-time task done during the enrollment phase by the server---the trusted authority. Most importantly, it costs no extra area and power overhead to the PUF integrated device, while it also avoids the overhead related to the ECC and helper data during both, key enrollment and regeneration phases. The challenge filtering procedure is described in the {\bf Algorithm.~\ref{Algorithm:SelectionReliableResponse}}.

\begin{algorithm}[h]
	\small
	\caption{Filtering challenges that generate reliable responses based on the statistical models of APUFs}
	\label{Algorithm:SelectionReliableResponse}
	\begin{algorithmic}[1]
		\Procedure{$\mathbf{selection}$~} {a randomly chosen challenge $ \bf C $, PUF model $ f(\cdot) $, $ \Delta t $}
		\State $ t_{\rm dif} \leftarrow f(\bf C)$
		
		\If {$ t_{\rm dif} < - \Delta t $}
		\State response $ \leftarrow $ 0; select $ \bf C $;
		\State \textbf{return}
		\ElsIf {$ t_{\rm dif} $ $ > $ $ \Delta t $}
		\State response $ \leftarrow $ 1; select $ \bf C $;
		\State \textbf{return}
		\Else
		\State  discard $ \bf C $;
		\State \textbf{return}
		\EndIf
		\EndProcedure
		\Statex
	\end{algorithmic}
	\vspace{-0.4cm}%
\end{algorithm}

\section{Experimental Validations}\label{Sec:ExperimentalSysthesis}
In this section, we empirically evaluate the reliable challenge selection method employing two datasets: i) we build up synthetic APUFs using ROPUFs~\cite{maiti2010dataset} frequency measurements, by this means, we are able to obtain almost arbitrary number of CRPs for extensively testings; ii) we use the CRP data collected across eight APUFs implemented on eight FPGAs~\cite{majzoobi2014automated}, the total number of CRPs in this dataset is 64,000, which is the main reason that we consider the synthesized APUFs first to obtain arbitrary number of CRPs due to 64,000 CRPs insufficient for some tests. For example, by using the first dataset, we test up to 50 million reliable challenges, all yielding error free responses, filtered from 830 million random challenges. This large number empirical tests is hard to be carried out based on the second dataset.

Building a statistical model of an APUF only requires collecting a number, 10,000, of CRPs under the nominal condition rather than all operating conditions. Time of CRP collection costs less than one second considering that one CRP evaluation needs 50 ns for a 64-stage APUF~\cite{lim2005extracting}. The delay characterization takes only less than fifteen seconds to obtain a model, where we use MATLAB 2012b to achieve the model building and the processor is an Intel i7-3770CPU@3.4GHz CPU.
\subsection{Dataset Descriptions}	
There are five ROPUFs implemented across five Spartan3E S500 FPGAs boards in Virginia Tech's public ROPUF data. Each FPGA implements one ROPUF that consists of 512 ring oscillators (ROs). Detailed implementation information is detailed in \cite{maiti2010large}. The dataset contains each RO's frequency measurements. Each RO's frequency is measured 100 times under 0.96~V, 1.08~V, 1.20~V, 1.32~V, 1.44~V, respectively, at a fixed temperature of $25\celsius $ to reflect supply voltage influence. Similarly, each RO is also evaluated 100 times under $35\celsius $, $45\celsius $, $55\celsius $, $65\celsius $, respectively, with a fixed supply voltage of 1.20~V, to reflect influence from temperature changes. To use these measurements for synthesizing an APUF, we employ the inverse of frequencies of four ROs to replace $ t_{13}^i, t_{24}^i, t_{14}^i, t_{23}^i  $ in order to form the realistic delay time behavior of the $ i $-th stage in the APUF. By this means, we are able to exploit existing real time delay data to form a $ 64 $-stage synthesized APUF by using 256 ROs implemented in the same FPGA board. The arbiter (performing delay time comparison) function is carried out off-chip by post-processing in MATLAB. Consequently, we obtain five synthesized APUFs. 

The second CRP dataset are obtained from eight PDL (programmable delay line) APUFs; each has 128 stages. Each APUF is fed with 64000 challenges, therefore, 64000 CRPs are collected~~\cite{majzoobi2009techniques,majzoobi2010fpga}. For each CRP, it is evaluated 128 times the same operating condition. In total, nine operating conditions are considered: ($ 5\celsius $, 0.95~V); ($ 5\celsius $, 1.00~V); ($ 5\celsius $, 1.05~V); ($ 35\celsius $, 0.95~V); ($ 35\celsius $, 1.00~V); ($ 35\celsius $, 1.05~V); ($ 65\celsius $, 0.95~V); ($ 65\celsius $, 1.05~V); ($ 65\celsius $, 1.05~V). We treat ($ 35\celsius $, 1.00~V) as the nominal condition.
\begin{figure}
\centering
\includegraphics[trim=0 0 0 0,clip,width=0.40\textwidth]{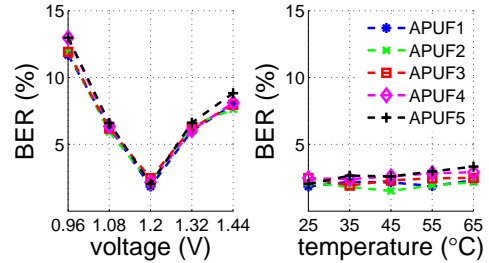}
\caption{Bit error rate (BER) under different supply voltage and temperature settings. Results are in well agreement with other experimentally reported results~\cite{roel2012physically,maes2012experimental}.}
\label{fig:BER}
\end{figure}

\subsection{Results of Synthesized APUFs}
We evaluate each APUF's bit error rate (BER). The BER is the probability that two PUF responses from two distinct and random evaluations subject to the same randomly chosen challenge applied to the same PUF are different. In practice, a reference response took under the nominal condition is always used, while the regenerated response took under a different operating condition is compared with the reference response~\cite{roel2012physically}. In our tests, each response bit is evaluated 100 times given the same challenge under the same operating condition. Results are shown in Fig.~\ref{fig:BER}. The BER solely introduced by noise is around $ 2.2\% $ when both voltage and temperature are under the nominal condition, where the voltage is 1.2~V and the temperature is $ 25\celsius $. We can see that supply voltage has predominant influence on the APUF BER compared with the temperature. Further, the worst-case BER of $ 12.98\% $ is observed when the voltage has a $ -20\% $ deviation from the nominal condition. Although APUFs evaluated here are synthesized, the presented reliability performance is in good agreement with~\cite{roel2012physically,maes2012experimental}. 

In the following descriptions, for convenience, we refer to a challenge that satisfies $ |t_{\rm dif}|>\Delta {t}  $ as a reliable challenge. Concurrently, we refer to the response bit corresponding to a reliable challenge as a reliable response. The statistical model has 97\% response accuracy prediction that is well agreed with~\cite{xu2015adaptive}. After reliable challenges are selected by performing the challenge filtering {\bf Algorithm~\ref{Algorithm:SelectionReliableResponse}}, they are applied to the same APUF when operating conditions are changed to any setting within a wide range. The regenerated response bits are compared with the reference response bit evaluated at the nominal condition. If they are same, this indicates that such a filtered challenge does have a very high tolerance to the environmental changes, which guarantees an error free response bit regenerated across a wide operating conditions. Otherwise, an error is marked and counted. We calculate the error rate that is the percentage of counted errors out of the number of selected reliable challenges. 
It is convenient to refer the error rate under a specific $ \Delta t $ setting as BER@$ \Delta t $. We denote the BER@$ \Delta t $ as below.

{\bf BER@$ \Delta t $.} The BER@$ \Delta t $ is the probability that the reevaluated response bit took under one random chosen operating condition within a range is different from the reference response bit evaluated under the nominal operating condition by applying the same selected reliable challenge to the same PUF, where the reliable challenge resulted delay time difference between top and bottom paths in an APUF satisfying $ |t_{\rm dif}|>\Delta {t} $.

We also refer BER@($ \Delta {t}=0 $) as BER@Default. In fact, BER@Default is the BER when challenge filtering is not employed, eg., shown in Fig.~\ref{fig:BER}.
\begin{table}
	\centering 
	\caption{BER@$ \Delta t$.~Synthetic APUF operational range: supply voltage(1.08-1.32~V), temperature(25$ \celsius $-65$ \celsius $)}
\resizebox{0.40\textwidth}{!}{%
	\begin{tabular}{c | c | c | c | c | c |c} 
		\toprule 
		\begin{tabular}{@{}c@{}} APUF\\No\end{tabular} & \begin{tabular}{@{}c@{}} BER@\\Default\end{tabular} & \begin{tabular}{@{}c@{}} BER@\\($ \Delta t$\\$=0.5 $)\end{tabular} & \begin{tabular}{@{}c@{}} BER@\\($ \Delta t$\\$=0.75 $)\end{tabular} &\begin{tabular}{@{}c@{}} BER@\\($ \Delta t$\\$=1.0 $)\end{tabular} & \begin{tabular}{@{}c@{}} BER@\\($ \Delta t$\\$=1.25 $)\end{tabular} & \begin{tabular}{@{}c@{}} BER@\\($ \Delta t$\\$=1.5 $)\end{tabular}\\ 
		\midrule 
		1 & 6.30\% & 0\% & 0\% & 0\% & 0\% & 0\%\\ 
		2 & 6.43\% & 0.013\% & 0\% & 0\% & 0\% & 0\%\\ 
		3 & 6.21\% & 0.00205\% & 0\% & 0\% & 0\% & 0\%\\ 
		4 & 6.34\% & 0.007\% & 0\% & 0\% & 0\% & 0\%\\ 
		5 & 6.60\% & 0.01\% & 0.0008\% & 0\% & 0\% & 0\%\\ 
		\bottomrule 
	\end{tabular}}
	\label{tab:1} 
\end{table}
\subsubsection{Reliability of Filtered Responses}
\begin{table}
	\centering 
	\caption{BER@$ \Delta t $.~Synthetic APUF operational range: supply voltage(0.96-1.44~V), temperature(25$ \celsius $-65$ \celsius $)}
\resizebox{0.40\textwidth}{!}{%
	\begin{tabular}{c | c | c | c | c | c |c} 
		\toprule 
		\begin{tabular}{@{}c@{}} APUF\\No\end{tabular} & \begin{tabular}{@{}c@{}} BER@\\Default\end{tabular} & \begin{tabular}{@{}c@{}} BER@\\($ \Delta t$\\$=0.5 $)\end{tabular} & \begin{tabular}{@{}c@{}} BER@\\($ \Delta t$\\$=0.75 $)\end{tabular} &\begin{tabular}{@{}c@{}} BER@\\($ \Delta t$\\$=1.0 $)\end{tabular} & \begin{tabular}{@{}c@{}} BER@\\($ \Delta t$\\$=1.25 $)\end{tabular} & \begin{tabular}{@{}c@{}} BER@\\($ \Delta t$\\$=1.5 $)\end{tabular}\\ 
		\midrule 
		1 & 11.68\% & 0.91\% & 0.1338\% & 0.0041\% & 0\% & 0\%\\ 
		2 & 11.90\% & 0.18\% & 0.13\% & 0.0056\% & 0.0006\% & 0\%\\ 
		3 & 11.92\% & 1.44\% & 0.22\% & 0.0023\% & 0.001\% & 0\%\\ 
		4 & 12.97\% & 1.29\% & 0.18\% & 0.0115\% & 0\% & 0\%\\ 
		5 & 12.98\% & 1.76\% & 0.34\% & 0.05\% & 0.0008\% & 0\%\\ 
		\bottomrule 
	\end{tabular}}
	\label{tab:2} 
\end{table}
Tested results are shown in Table.~\ref{tab:1} and~\ref{tab:2}. The BER@$ \Delta t $ is evaluated within different operating ranges for each Table.  
Noting the supply voltage range of Table.~\ref{tab:2} ($ \pm 20\% $ deviation) is larger than Table.~\ref{tab:1} ($ \pm 10\% $ deviation). 
It is not surprising that BER@$ \Delta t $ is not zero when the $ \Delta t $ is set to a small value. But as the $ \Delta t $ goes up, BER@$ \Delta t $ decreases significantly. When the $ \Delta t=1.5 $, all filtered challenges that satisfy $ |t_{\rm dif}|>(\Delta t=1.5) $ produce responses with 100\% reliability. The APUF$ _5 $ is the most interested testing sample because, in Fig.~\ref{fig:BER}, APUF$ _5 $ has the highest worst-case BER of 12.98\% among other APUFs when the voltage varies between 0.96 V and 1.44 V. By using judiciously selected error free responses to derive keys, it equals to an ECC decoder that has to correct at least 12.98\% errors. 

Due to infinite number of reliable challenges filtering cannot be achieved in evaluation, we use up to more than 830 million randomly generated challenges for all testings, among them, 50 million selected reliable challenges are obtained when the $ \Delta t=1.5 $. We can see that all of responses given selected challenges are error free regenerated even when the supply voltage has a $ \pm 20\% $ variation and the temperature experiences a $ 40\celsius $ variation.
\subsubsection{CRP Loss}
\begin{figure}
	\centering
	\includegraphics[trim=0 0 0 0,clip,width=0.30\textwidth]{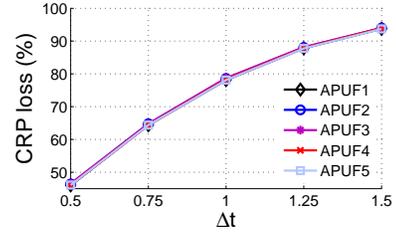}
	\caption{Illustration of CRP loss induced by the reliable challenge filtering.}
	\label{fig:bitloss} 
\end{figure}
The CRP loss is the probability that a randomly given challenge cannot meet the challenge filtering criterion of $ |t_{\rm dif}|>\Delta t  $. The CRP loss is increasing as the $ \Delta t  $ increases. Fig.~\ref{fig:bitloss} depicts the CRP loss as a function of $ \Delta t  $. We can see that 94\% response bits will be discarded when $ |t_{\rm dif}|>(\Delta t=1.5)  $. As an APUF is able to generate an exponential number of CRPs as a function of the number of stages $ k $, large number of reliable challenges can still be promised. For example, given a 64-stage APUF, its CRP space is up to $ 2^{64} $. Hence, there are still $ 2^{64}\times 0.06>2^{59} $ error free response bits available. When the PUF operating range varies insignificant in practice, a smaller $ \Delta t  $ can be chosen to decrease the CRP loss while still produce error free responses as shown in Table.~\ref{tab:1}. 
\subsection{PDL APUFs}
\begin{table}
		\centering 
		\caption{BER@$ \Delta t $. PDL APUF operational range: supply voltage(0.95-1.05~V), temperature(5$ \celsius $-65$ \celsius $)}
\resizebox{0.40\textwidth}{!}{%
		\begin{tabular}{c | c | c | c | c | c |c} 
			\toprule 
		\begin{tabular}{@{}c@{}} APUF\\No\end{tabular} & \begin{tabular}{@{}c@{}} BER@\\Default\end{tabular} & \begin{tabular}{@{}c@{}} BER@\\($ \Delta t$\\$=0.5 $)\end{tabular} & \begin{tabular}{@{}c@{}} BER@\\($ \Delta t$\\$=1.0 $)\end{tabular} &\begin{tabular}{@{}c@{}} BER@\\($ \Delta t$\\$=1.5 $)\end{tabular} & \begin{tabular}{@{}c@{}} BER@\\($ \Delta t$\\$=1.6 $)\end{tabular} & \begin{tabular}{@{}c@{}} BER@\\($ \Delta t$\\$=1.7 $)\end{tabular}\\ 
			\midrule 
			1 & 10.57\% & 1.26\% & 0.095\% & 0\% & 0\% & 0\%\\ 
			2 & 16.68\% & 8.05\% & 1.46\% & 0.039\% & 0\% & 0\%\\ 
			3 & 8.59\% & 1.08\% & 0.037\% & 0\% & 0\% & 0\%\\ 
			4 & 8.62\% & 0.656\% & 0.013\% & 0\% & 0\% & 0\%\\ 
			5 & 11.52\% & 0.08\% & 0\% & 0\% & 0\% & 0\%\\ 
			6 & 11.07\% & 0.71\% & 0.031\% & 0\% & 0\% & 0\%\\ 
			7 & 5.87\% & 1.35\% & 0.032\% & 0\% & 0\% & 0\%\\ 
			8 & 15.00\% & 5.98\% & 0.89\% & 0.056\% & 0.085\% & 0\%\\ 
			\bottomrule 
		\end{tabular}}
		\label{tab:PDLAPUFs} 
	\end{table}
The statistical model is obtained by training 10,000 CRPs evaluated only under the nominal condition and the prediction accuracy with 92.41\% is achieved. This lower prediction accuracy compared with the prediction accuracy of the synthetic APUF model are attributed to~\cite{ruhrmair2013pufs}: i) The slightly increased nominal BER is 4.99\% that is obtained under the nominal condition---the nominal BER of the synthesized APUF is 2.2\%, see Fig.~\ref{fig:BER}; ii) The usage of PDLs on FPGA, which makes the MUX structure more complicated (slightly nonlinearity is introduced).

There are eight PDL APUFs tested, each of them is implemented on a different FPGA board. The maximum BER@$ \Delta t $ for all of eight APUFs are listed in Table.~\ref{tab:PDLAPUFs}. Among tested eight APUFs, the maximum worst-case BER@Default is 16.68\%. When the $ \Delta t=1.7 $, there is no error found in all reproduced responses under any tested operating condition. Same to the results obtained from synthesized APUFs in \ref{tab:2}, BER@$ \Delta t$ decreases as the $ \Delta t $ increases. There are still around 1\% challenges satisfying the selection criterion for all eight tested APUFs even when $ \Delta t=1.7 $.
\begin{figure}
	\centering
	\includegraphics[trim=0 0 0 0,clip,width=0.35\textwidth]{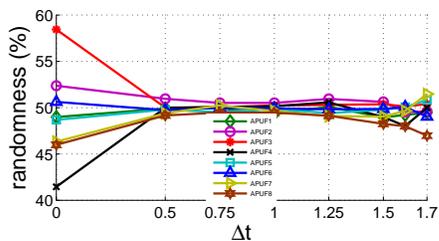}
	\caption{Randomness under different $ \Delta t $ settings.}
	\label{fig:randomness} 
\end{figure}

We further investigate the response randomness---percentage of occurrence of `1's in response bits---relationship with the $ \Delta t $, In general, the challenge filtering should not result into a bias to PUF's response with a preferable value (`0'/`1'). In Fig.~\ref{fig:randomness}, the challenge filtering does not deteriorate the randomness of the selected reliable response bits. The randomness is always close to 50\% when the challenge filtering is performed under different $ \Delta t $ settings. In addition, even the response's randomness is not close to 50\% initially, the challenge filtering seems eliminate such an initial randomness bias. 
\section{Conclusion}\label{Sec:Conclusion}
We propose an approach for determining error free responses even when they are re-evaluated across a wide range of operating condition. The proposed approach can be adopted into controlled PUF designs and also provide large number of error free responses on demand for advanced cryptographic applications. This method has no burden to the PUF embedded devices because there is no ECC and helper data involved, where the model building is left to the server and only asks negligible computational resources to the server. Extensive empirically evaluations validate the practicability of our error free response methodology. There are some interesting future works. Firstly, though good results have been demonstrated through simulated data~\cite{xuusing} considering aging effects, further empirical validations remain to be investigated. Secondly, adopting our error-free response generation method in to controlled PUFs~\cite{gassend2008controlled}. In this context, it is imperative to reduce the CRP loss, which will enable an efficient control logic realization within the controlled PUFs. Last but not least, attacks assisted by the helper data compromising the security of a key generator only extracting a single key or very limited number of keys~\cite{becker2015pitfalls} may be even harder, if not impossible, when they are mounted to key generators in capable of generating a large number of keys where the helper data is not used. This will be another interesting investigation of securely using error-free responses~\cite{gao2017puf}.
\section{Acknowledgment}
We acknowledge the APUF dataset provided by Dr Mehrdad Majzoobi and Mr Siam U. Hussain from research group of Prof Farinaz Koushanfar.
%

\end{document}